\documentclass{amsart}
\usepackage{amsmath,amssymb}
\usepackage{hyperref}
\usepackage{mathrsfs}  

\newtheorem{theo}{Theorem}[section]

\newcommand{\hgm}{\hat{\mathfrak{g}}}
\newcommand{\iN}{\lrcorner}
\newtheorem{theorem}{Theorem}[section]
\newtheorem{lemma}[theorem]{Lemma}

\begin{document}

\title{A variational principle for Kaluza--Klein type theories}


\author{Fr{\'e}d{\'e}ric H{\'e}lein}
\address{IMJ-PRG,
UMR CNRS 7586 \& Universit{\'e} Denis Diderot,
UFR de Math{\'e}matiques,  B{\^a}timent Sophie Germain,
8 place Aur{\'e}lie Nemours, 75013 Paris, France}
\curraddr{}
\email{helein@math.univ-paris-diderot.fr}
\thanks{}



\date{October 25, 2018}

\dedicatory{}

\begin{abstract}
For any positive integer $n$ and any Lie group $\mathfrak{G}$,
given a definite symmetric bilinear form on $\mathbb{R}^n$ and an $\hbox{Ad}$-invariant scalar product
on the Lie algebra of $\mathfrak{G}$,
we construct a variational problem on fields defined on an arbitrary oriented
$(n+\hbox{dim}\mathfrak{G})$-dimensional manifold $\mathcal{Y}$.
We show that, if $\mathfrak{G}$ is compact and simply connected, any global solution
of the Euler--Lagrange equations leads, through a spontaneous symmetry breaking,
to identify $\mathcal{Y}$ with the total space of a principal bundle
over an $n$-dimensional manifold $\mathcal{X}$. Moreover $\mathcal{X}$
is then endowed with a (pseudo-)Riemannian metric and a connection which are solutions
of the Einstein--Yang--Mills system of equations with a cosmological constant.
\end{abstract}

\maketitle

\tableofcontents

\section{Introduction}
\subsection{Motivations}
In 1919 T. Kaluza \cite{kaluza} (after an earlier attempt by G. Nordstr{\"o}m \cite{nordstrom} in 1914) discovered
that solutions of the Einstein equations of gravity in vacuum on a 5-dimensional manifold
could modelize Einstein equations coupled with Maxwell equations on a 4-dimensional 
space-time manifold, provided one assumes that the 5-dimensional manifold is a circle
fiber bundle over space-time and that the metric is constant along these fibers.
This was rediscovered more or less independentely by O. Klein \cite{klein} in 1926 (and also by
H. Mandel \cite{mandel}), who proposed to assume that the size of the extra fifth dimension is sufficientely
tiny in order to explain why this dimension is not directly observed. Since then this fascinating observation
has been an important source of inspiration and questioning (see e.g. \cite{einsteinbergmann}). It has been extended
to include non Abelian gauge theories \cite{dewitt,kerner,chofreund,cremmer-scherk}, in
order to unify the Einstein equations
with the Yang--Mills equations on a curved space-time and, in particular, it becomes
an important ingredient of the 11-dimensional supergravity and the superstrings theories.
It remains today a subject of questioning (see e.g. \cite{appelquist,witten}).\\

However some difficulties plag this beautiful idea:

The Kaluza--Klein ansatz is indeed based on the assumption that the metric
is covariantly constant along the fibers. But this raises the question of
finding physical reasons for that. Moreover
the initial proposal by Kaluza and Klein led 
to inconsistency. This point was raised by P. Jordan \cite{jordan} and Y. Thiry \cite{thiry},
who allowed the coefficient of the metric along the fifth dimension to be an extra scalar field.
However this scalar field is a source of difficulties as to its physical interpretation.

A way to avoid the assumption that the metric is covariantly constant along
the fibers is, as proposed by Klein, to assume that the extra dimension is tiny.
Then by expanding the fields in harmonic modes on each fiber one finds
that, as a consequence of the Heisenberg uncertainty principle,
all modes excepted the zero one should be extremely massive.
This would hence explain why we cannot observe their quantum excitations.
This idea is at the origin of the current hypothesis.

But this does not answer the fundamental question of understanding why
these extra dimensions are fibered and compact (and tiny if we want to support the above hypothesis or, alternatively, if the smallness assumption is not true,
why the metric is constant along the fiber):
could a dynamical mechanism explain these assumptions~?\\

In the following we address these questions and we
present a variational principle which satisfies the following
properties: provided that the involved structure Lie group is compact and simply connected,
the Euler--Lagrange equations satisfied by the critical points
lead to a mechanism which forces a spontaneous
fibration of the higher dimensional manifold over an emerging space-time, forces the 
metric to be covariantly constant along the fibers and one can 
build out of these critical points a metric and a connection over the
space-time which are solutions
of the Einstein--Yang--Mills system of equations.

Note that our results work partially for e.g. $U(1)$, for which our mechanism fails
to imply the compactness of the fibers
without extra \emph{ad hoc} hypotheses. Hence either there is a need to improve
our theory (for example by taking into account semi-classical or quantum effects),
or one may argue that our results could be sufficient in an Grand Unified Theory,
where all structure gauge groups are supposed to arise from
a single compact, simply connected one, by a symmetry breaking.

\subsection{The main result}
To introduce our model let us first remind the higher dimension generalization
of the so-called \emph{Palatini} (see \cite{ferraris}) formulation of gravity:
Let $N\geq 2$ be an integer and $E$ be an oriented $N$-dimensional real vector space
endowed with a non degenerate bilinear form $\textsf{h}$ (in most cases the
Minkowski scalar product) and let $so(E,\textsf{h})$ be the Lie algebra
of the group of isometries of $(E,\textsf{h})$. We identify
$so(E,\textsf{h})$ with $\Lambda^2 E = E\wedge E$ (with a Lie bracket
denoted by $[\cdot,\cdot]_2$, see the next
section for details).
The $N$-dimensional generalization of the Palatini
action on an oriented $N$-dimensional manifold $\mathcal{Y}$
is a functional defined on pairs $(\theta,\varphi)$ where
$\theta$ is a (soldering) 1-form on $\mathcal{Y}$ with coefficient in $E$
and $\varphi$ is a (connection) 1-form on
$\mathcal{Y}$ with coefficient in $so(E,\textsf{h})\simeq \Lambda^2 E$.
This functional reads
\[
 \mathscr{A}_P[\theta,\varphi] =
 \int_\mathcal{Y} \star\theta^{(N-2)}\wedge 
 \left(d\varphi + \frac{1}{2}[\varphi\wedge \varphi]_2\right),
\]
or $\mathscr{A}_P[\theta,\varphi] = \int_\mathcal{Y} \star\theta^{(N-2)}\wedge \Phi$
by denoting $\Phi:= d\varphi + \frac{1}{2}[\varphi\wedge \varphi]_2$.
Here $\star\theta^{(N-2)}$ is the $(N-2)$-form with coefficient in $so(E,\textsf{h})^*\simeq \Lambda^2 E^*$,
with components $\theta^{(N-2)}_{A_1A_2} = 
 \frac{1}{(N-2)!} \epsilon_{A_1\cdots A_N}
 \theta^{A_3}\wedge \cdots \wedge \theta^{A_N}$,
where $\epsilon_{A_1\cdots A_N}$ is the completely antisymmetric
tensor such that $\epsilon_{1\cdots N} = 1$ and,
in the product $\star\theta^{(N-2)}\wedge \Phi$, the duality pairing 
between $\Lambda^2 E^*$ and $\Lambda^2 E$ is implicitely assumed
so that
$\star\theta^{(N-2)}\wedge \Phi
= \frac{1}{2} \theta^{(N-2)}_{AB} \wedge \Phi^{AB}$ (see the next section for more details).

Then, as it is well-known, the critical points $(\theta,\varphi)$ of $\mathscr{A}_P$ such that the 
rank of $\theta$ is equal to $N$ everywhere correspond to solutions of the 
Einstein equations of gravity in vacuum (with a metric $\theta^*\textsf{h}$
on $\mathcal{Y}$).\\

Our model can be seen as a deformation of the previous one:
we assume that $E$ is itself endowed with a Lie bracket $[\cdot,\cdot]_1$
and we denote by $\hgm : =(E,[\cdot,\cdot]_1)$ the resulting Lie algebra.
We assume further that:
\begin{enumerate}
 \item[(i)] $\hgm = \mathfrak{s}\oplus \mathfrak{g}$, where $\mathfrak{s}$
 is contained in the center of $\hgm$ and $\mathfrak{g}$ is a Lie subalgebra;
\item[(ii)] the Lie bracket
$[\cdot,\cdot]_1: \hgm\times \hgm \longrightarrow \hgm$
preserves the metric $\textsf{h}$;
\item[(iii)] $\mathfrak{s}$ is orthogonal to $\mathfrak{g}$ for the 
bilinear form $\textsf{h}$.
\end{enumerate}
We set $n = \hbox{dim}\mathfrak{s}$ and $r = \hbox{dim}\mathfrak{g}$
so that $N=n+r$. 
Note that (i) implies that $\mathfrak{s}$ is a \emph{trivial} Lie subalgebra
and (ii) means that 
$\forall \xi\in \hgm$, $\hbox{ad}_\xi\in so(\hgm,\textsf{h})$.
We consider the following space of fields: 
\[
\begin{array}{rlr}
 \mathscr{E}:=
 \{\, (\theta,\varphi,\pi); & \theta\in \hgm \otimes\Omega^1(\mathcal{Y}),\;
  \varphi\in so(\hgm,\textsf{h}) \otimes\Omega^1(\mathcal{Y}), & \\
  & \pi \in  \hgm^*\otimes\Omega^{N-2}(\mathcal{Y})
 &  \}
\end{array}
\] 
and define on it the action functional $\mathscr{A}$ by:
\begin{equation}
 \mathscr{A}[\theta,\varphi,\pi]:= \int_{\mathcal{Y}}  \pi \wedge \left(d\theta
 + \frac{1}{2}[\theta\wedge \theta]_1\right)
 + \star \theta^{(N-2)} \wedge \left(d\varphi + \frac{1}{2}[\varphi\wedge \varphi]_2\right)
\end{equation}
where the duality pairing between, respectively, $\hgm^*$ and $\hgm$ and
$so(\hgm,\textsf{h})^*$ and $so(\hgm,\textsf{h})$ is implicitely
used.

We decompose $\theta = \theta^\mathfrak{s} + \theta^\mathfrak{g}$
according to the splitting  $\hgm = \mathfrak{s}\oplus \mathfrak{g}$ and
we impose the constraint
\begin{equation}\label{deuxiemecontrainte}
 \theta^\mathfrak{s}\wedge \theta^\mathfrak{s} \wedge \pi =0
\end{equation}
(see the next section for more details)
leading hence us to define the constrained subset:
\[
 \mathscr{C}:= \{ (\theta,\varphi,\pi)\in\mathscr{E};\quad \theta^\mathfrak{s}\wedge \theta^\mathfrak{s} \wedge \pi =0 \}.
\]
\begin{theo}
Assume Hypotheses (i), (ii), (iii).
Let $\mathcal{Y}$ be a connected, oriented $N$-dimensional manifold.
Let $(\theta,\varphi,\pi)\in \mathscr{C}$ be a smooth critical point of the restriction
of $\mathscr{A}$
 on $\mathscr{C}$. Let $\textbf{h}:= \theta^*\textsf{h}$, a
 pseudo Riemannian metric on $\mathcal{Y}$.
 Assume that:
 \begin{enumerate}
  \item[(iv)] $\mathfrak{g}$ is the Lie algebra of a
  \textbf{compact} and \textbf{simply connected} Lie group $\mathfrak{G}$;
 \item[(v)] the rank of $\theta$ is equal to $N$ everywhere;
 \item[(vi)] $\textbf{h}:= \theta^*\textsf{h}$ is \emph{vertically complete} (see \S \ref{about}).
 \end{enumerate}
Then
 \begin{enumerate}
  \item the manifold $\mathcal{Y}$ is the total space of a principal bundle over
  an $n$-dimensional manifold $\mathcal{X}$;
  \item the structure group of this bundle is a group $\mathfrak{G}_0$, the universal cover of
  which is $\mathfrak{G}$;
  \item we can construct explicitely out of $\theta$ a pseudo Riemannian metric $\textbf{g}$
  and a $\mathfrak{g}$-valued connection $\nabla$ on $\mathcal{X}$;
  \item $\textbf{g}$ and $\nabla$ are solution of the Einstein--Yang--Mills system with
  cosmological constant equal to $\Lambda = \frac{1}{8}(K,\textsf{h}^*)$,
  where $K$ is the Killing form on $\hgm$, $\textsf{h}^*$ is the metric on
  ${\hgm}^*$ and $(\cdot,\cdot)$ is the natural pairing between both tensors.
 \end{enumerate}
\end{theo}\label{theorem0}
\subsubsection{About Hypothesis (vi)}\label{about}
The pseudo Riemannian metric $\textbf{h}:= \theta^*\textsf{h}$
 is \emph{vertically complete} if, for any continuous
map $v$ from $[0,1]$ to $\mathfrak{g}\subset \hgm$ and, for any
point $\textsf{y}\in \mathcal{Y}$, there exists 
an unique $\mathscr{C}^1$ map $\gamma:[0,1]\longrightarrow \mathcal{Y}$,
which is a solution of $(\gamma^*\theta)_t= v(t)dt$, $\forall t\in [0,1]$, with the
initial condition $\gamma(0) = \textsf{y}$.
Such curves $\gamma$ can be interpreted
\emph{a posteriori} as being \emph{vertical} curves, i.e.
contained in a fiber of the principal bundle over 
a point in the space-time. This allows thus singular space-times with black holes.

\subsubsection{Remark}
Our action may alternatively be written as follows.
We endow the direct sum $\hgm\oplus so(\hgm,\textsf{h})$
with the product Lie bracket $[\cdot,\cdot]$ of, respectively, $(\hgm,[\cdot,\cdot]_1)$ and
$(so(\hgm,\textsf{h}),[\cdot,\cdot]_2)$. We consider the space of fields
\[
\begin{array}{rll}
 \tilde{\mathscr{E}}:=
 \{\, (\theta+\varphi,\pi+\psi); & \theta\in \hgm \otimes\Omega^1(\mathcal{Y}),
 & \varphi\in so(\hgm,\textsf{h}) \otimes\Omega^1(\mathcal{Y}),\\
  & \pi \in  \hgm^*\otimes\Omega^{N-2}(\mathcal{Y}),
 & \psi \in  so(\hgm,\textsf{h})^*\otimes\Omega^{N-2}(\mathcal{Y})
 \}
\end{array}
\]
and we define
\[
 \tilde{\mathscr{A}}[\theta+\varphi,\pi+\psi]:= \int_{\mathcal{Y}}  (\pi+\psi) \wedge \left(d(\theta+\varphi)
 + \frac{1}{2}[\theta+\varphi\wedge \theta+\varphi]\right).
\]
We observe that, if the constraint
\begin{equation}\label{contraintebrute}
 \psi = \star\theta^{(N-2)}
\end{equation}
is satisfied, then $\tilde{\mathscr{A}}[\theta+\varphi,\pi+\psi] = \mathscr{A}[\theta,\varphi,\pi]$.
Hence the study of critical points of $\mathscr{A}$ on $\mathscr{C}$ is equivalent to the study
of critical points of $\tilde{\mathscr{A}}$ on:
\[
 \tilde{\mathscr{C}}:= \{ (\theta+\varphi,\pi+\psi)\in\tilde{\mathscr{E}};
 \quad \psi = \star\theta^{(N-2)} \hbox{ and } \theta^\mathfrak{s}\wedge \theta^\mathfrak{s} \wedge \pi =0 \}.
\]


\subsection{Outline of the proof} --- 
The action $\mathscr{A}[\theta,\varphi,\pi]$ is the sum of the generalized Palatini
action $\int_\mathcal{Y}\star\theta^{(N-2)}\wedge (d\varphi+\frac{1}{2}[\varphi\wedge \varphi]_2)$
and of the extra term $\int_\mathcal{Y}\pi\wedge (d\theta+\frac{1}{2}[\theta\wedge\theta]_1)$.

In the latter term the coefficients of the $(N-2)$-form $\pi$ (constrained
by $\theta^\mathfrak{s}\wedge \theta^\mathfrak{s}\wedge \pi = 0$)
play the role of Lagrange multipliers and, for a critical point,
it forces $d\theta+\frac{1}{2}[\theta\wedge\theta]_1$
to be a linear combination of components of $\theta^\mathfrak{s}\wedge \theta^\mathfrak{s}$.
One can thus use repeatedly Frobenius theorem: first to the Pfaffian system
$\theta^\mathfrak{s}|_{\textsf{f}} = 0$, where $\textsf{f}$ is an $r$-dimensional
submanifold of $\mathcal{Y}$, to obtain a local foliation of $\mathcal{Y}$, the leaves $\textsf{f}$
that we show are actually the fibers of a fibration $\mathcal{Y}\longrightarrow \mathcal{X}$
thanks to the hypotheses; second by using the fact that
$d\theta^\mathfrak{g}+\frac{1}{2}[\theta^\mathfrak{g}\wedge\theta^\mathfrak{g}]_1$
is a linear combination of components of $\theta^\mathfrak{s}\wedge \theta^\mathfrak{s}$
to deduce that the geometric data associated with $\theta$ are covariantly constant
along the fibers.

On the other hand one uses the fact that the first variation of $\mathscr{A}$ with respect to $\varphi$ vanishes
to show that the connection on $T\mathcal{Y}$ associated to $\varphi$ and $\theta$ is
the Levi-Civita connection for the metric $\textbf{h}:= \theta^*\textsf{h}$.
Note that this step is the same as in the standard derivation of the Palatini
Euler--Lagrange equation since $\varphi$ is only present in the integral
$\int_\mathcal{Y}\star\theta^{(N-2)}\wedge (d\varphi+\frac{1}{2}[\varphi\wedge \varphi]_2)$.

Lastly one exploits the fact that the first variation of $\mathscr{A}$ with respect to $\theta$ vanishes.
If the action would only be equal to
$\int_\mathcal{Y}\star\theta^{(N-2)}\wedge (d\varphi+\frac{1}{2}[\varphi\wedge \varphi]_2)$
one would find that the metric $\textbf{h}$ on $\mathcal{Y}$ is a solution of the Einstein equation
in vacuum and consequently the equivariance of the metric along the fibers derived previously
would then give us a solution of an Einstein--Yang--Mills system of equations on $\mathcal{X}$.
However the coupling of $\theta$ with $\pi$ in the second term
$\int_\mathcal{Y}\pi\wedge (d\theta+\frac{1}{2}[\theta\wedge\theta]_1)$
creates extra source terms in the Einstein--Yang--Mills system
which contains an \emph{a priori} high degree of arbitrariness
and which could
hence ruin our efforts.

\emph{A miraculous cancellation}: however,
apart from a cosmological constant,
the extra sources just cancel!
This cancellation is due to the fact that each of the source terms is
covariantly constant along each fiber and hence is equal to its
average value on the fiber, which is compact. But it turns out that
this average value is proportional to the integral of an \emph{exact} $r$-form on the fiber
and hence vanishes. This phenomenon is similar to the one discovered in \cite{helein14} and \cite{heleinvey15}.

\section{Notations and description of the obtained equations}
\subsection{Intrinsic setting}
Since our action and the resulting Euler--Lagrange equations mix forms with coefficients in
$\hgm$, $so(\hgm,\textsf{h})$ and their dual spaces it will be convenient to identify
$so(\hgm,\textsf{h})$ wih $\hgm\wedge \hgm$ as follows.

For any finite dimensional real vector space $E$ and any $k\in \mathbb{N}$ we let
$E^{\otimes k}$ be the $k$-th tensorial power of $E$  and 
$\Lambda^kE:= E\wedge \cdots \wedge E$ be the subspace of $E^{\otimes k}$ of skewsymmetric
tensors. If $v_1,\cdots, v_k\in E$ we set
\[
 v_1\wedge \cdots \wedge v_k:= \sum_{\sigma\in \mathfrak{S}(k)} (-1)^{|\sigma|}
 v_{\sigma(1)}\otimes \cdots \otimes v_{\sigma(k)}\in \Lambda^kE
\]
and, for $p\in \mathbb{N}$ greater than or equal to $k$ and $\lambda\in \Lambda^pE^*$,
we define the \emph{interior product}\footnote{Note that, if we view $v_1\wedge \cdots \wedge v_k$ and $\lambda$
as elements of, respectively, $E^{\otimes k}$ and $(E^*)^{\otimes p}$, then
$v_1\wedge \cdots \wedge v_k\iN \lambda$ is $1/k!$ times the contraction of 
$v_1\wedge \cdots \wedge v_k$ with $\lambda$.} $v_1\wedge \cdots \wedge v_k\iN \lambda$
to be the $(p-k)$-form in $\Lambda^{p-k}E^*$
such that:\\
$\left(v_1\wedge \cdots \wedge v_k\iN \lambda\right)(w_{k+1},\cdots ,w_p) = 
\lambda(v_1,\cdots, v_k,w_{k+1},\cdots ,w_p)$,
$\forall ,w_{k+1},\cdots ,w_p\in E$.


To any $\xi\otimes \alpha$ in $E\otimes E^*$ we associate the linear map from $E$ to itself
defined by $[\eta\longmapsto \xi\alpha(\eta)]$.
By extending linearly this map, we get a linear isomorphism which allows
us to identify $E \otimes E^*$ with $End(E)$.
If furthermore $E$ is endowed with 
a non degenerate symmetric bilinear form $\textsf{h}$, it induces a vector space 
isomorphism $\zeta\longmapsto \zeta\iN \textsf{h}:= \textsf{h}(\zeta,\cdot)$
from $E$ to $E^*$. We hence get an unique linear map $\mathscr{L}:E \otimes E \longrightarrow E \otimes E^*\simeq End(E)$
such that, for any $\xi,\zeta,\eta\in E$,
\[
\mathscr{L}(\xi\otimes \zeta) = \xi\otimes (\zeta\iN \textsf{h}) \simeq [E \ni \eta \longmapsto 
\xi\textsf{h}(\zeta,\eta)\in E].
\]
Then $\mathscr{L}$ is an isomorphism. We endow $E\otimes E$ with the unique product
law $\ast$ such
that $\mathscr{L}( \alpha\ast \beta) = \mathscr{L}(\alpha)\circ \mathscr{L}(\beta)$, 
$\forall \alpha,\beta\in E\otimes E$. We also get a Lie algebra bracket $[\cdot,\cdot]_2$
on $E\otimes E$ defined by $[\alpha, \beta]_2 = \alpha\ast \beta -\beta \ast \alpha$.

The subspace $\Lambda^2E=E\wedge E\subset E\otimes E$ 
is then a Lie subalgebra of $(E\otimes E, [\cdot,\cdot]_2)$
which coincides with the inverse image by $\mathscr{L}$
of the Lie subalgebra $so(E,\textsf{h})$.
This allows us to identify $so(E,\textsf{h})$ with $\Lambda^2E$ endowed
with the bracket $[\cdot,\cdot]_2$.


\subsection{Introducing a basis of $\hgm$ and using indices}
We let $(\textbf{t}_A)_{1\leq A\leq N}$ be a basis of $\hgm$ such that
$(\textbf{t}_a)_{1\leq a\leq n}$ is a basis of $\mathfrak{s}$ and 
$(\textbf{t}_\alpha)_{n+1\leq \alpha\leq N}$ is a basis of $\mathfrak{g}$.
We will systematically use the following conventions for the indices: $1\leq A,B,C,\ldots\leq N$
and 
$1\leq a,b,c,\ldots\leq n< \alpha,\beta,\gamma,\ldots\leq N$.

We denote by $\textsf{b}$ the restriction of $\textsf{h}$ to $\mathfrak{s}$
and $\textsf{k}$ the restriction of $\textsf{h}$ to $\mathfrak{k}$ and we set
$\textsf{h}_{AB}:= \textsf{h}(\textbf{t}_A,\textbf{t}_B)$,
$\textsf{b}_{ab}:= \textsf{b}(\textbf{t}_a,\textbf{t}_b)$ and
$\textsf{k}_{\alpha\beta}:= \textsf{k}(\textbf{t}_\alpha,\textbf{t}_\beta)$, so that
Hypothesis (iii) translates as
\[
\left(\textsf{h}_{AB}\right) 
= \left(\begin{array}{cc}
\textsf{h}_{ab} & \textsf{h}_{a\beta} \\ \textsf{h}_{\alpha b} & \textsf{h}_{\alpha\beta}
\end{array}\right)
= \left(\begin{array}{cc} \textsf{b}_{ab} & 0 \\ 0 & \textsf{k}_{\alpha\beta}
\end{array}\right).
\]           
We denote by $c^A_{BC}$ the structure constants of $\hgm$ in the basis $(\textbf{t}_A)_{1\leq A\leq N}$,
defined by $[\textbf{t}_B,\textbf{t}_C]_1 = \textbf{t}_Ac^A_{BC}$.
We observe that, due to Hypothesis (i),
\[
 \left(c^A_{BC}\right)
 = \left(\begin{array}{ccc}
          c^a_{bc} & c^a_{b\gamma } & c^a_{\beta \gamma} \\
          c^\alpha_{bc} & c^\alpha_{b\gamma } & c^\alpha_{\beta \gamma}
         \end{array}\right)
 = \left(\begin{array}{ccc}
          0 & 0 & 0 \\
          0 & 0 & c^\alpha_{\beta \gamma}
         \end{array}\right).
\]
For any $A,B = 1,\cdots,N$,
we let $\textbf{t}_{AB}:= \textbf{t}_A\wedge \textbf{t}_B$. Then
$\left(\textbf{t}_{AB}\right)_{1\leq A<B\leq N}$ is a basis of $\Lambda^2\hgm = \hgm\wedge \hgm$.
Hence using the isomorphism $\mathscr{L}$ defined previously to identify $so(\hgm,\textsf{h})$
with $\Lambda^2\hgm$ we can view
$\left(\textbf{t}_{AB}\right)_{1\leq A<B\leq N}$ as a basis of $so(\hgm,\textsf{h})$
as well. Through this identification we have
$\textbf{t}_{AB}(\textbf{t}_C) = \textbf{t}_A\textsf{h}_{BC} - \textbf{t}_B\textsf{h}_{AC}$.
Moreover
\[
 [\textbf{t}_{A_1B_1}, \textbf{t}_{A_2B_2}]_2
 = \textbf{t}_{A_1B_2}\textsf{h}_{B_1A_2}
 - \textbf{t}_{A_1A_2}\textsf{h}_{B_1B_2}
 - \textbf{t}_{B_1B_2}\textsf{h}_{A_1A_2}
 + \textbf{t}_{B_1A_2}\textsf{h}_{A_1B_2}.
\]
We denote by $(\textbf{t}^A)_{1\leq A\leq N}$ the basis of $\hgm^*$ which is dual to 
$(\textbf{t}_A)_{1\leq A\leq N}$ and by
$\left(\textbf{t}^{AB}\right)_{1\leq A<B\leq N}$ the basis of $\Lambda^2\hgm^*\simeq (\Lambda^2\hgm)^*$
which is dual to $\left(\textbf{t}_{AB}\right)_{1\leq A<B\leq N}$.

If $\Phi$ is a form with coefficients in $so(\hgm,\textsf{h})$ with coordinates
$\left(\Phi^{AB}\right)_{1\leq A<B\leq N}$ we set
$\Phi^{BA}:= - \Phi^{AB}$, for $A\geq B$, so that
\[
 \Phi = \sum_{1\leq A<B\leq N}\textbf{t}_{AB}\Phi^{AB}
 = \frac{1}{2}\sum_{A,B=1}^N\textbf{t}_{AB}\Phi^{AB}
 = \frac{1}{2}\textbf{t}_{AB}\Phi^{AB}
\]
and we will systematically use the last writing $\Phi = \frac{1}{2}\textbf{t}_{AB}\Phi^{AB}$,
where the summation over $1\leq A,B\leq N$ is implicitely assumed.
Similarly if $\psi$ is a $\Lambda^2\hgm^*$-valued form, we will use the same convention
$\psi = \frac{1}{2}\psi_{AB}\textbf{t}^{AB}$ for its decomposition in the basis
$\left(\textbf{t}^{AB}\right)_{1\leq A<B\leq N}$. The duality pairing between a
$\Lambda^2\hgm$-valued form $\Phi$ and a $\Lambda^2\hgm^*$-valued form
$\psi$ then reads $\psi\wedge \Phi = \frac{1}{2}\psi_{AB}\wedge \Phi^{AB}$.

Lastly we use $\textsf{h}_{AB}$ and $\textsf{h}^{AB}$ to rise and lower the indices:
$\varphi{^A}_B = \textsf{h}_{BB'}\varphi^{AB'}$, 
$\varphi^{AB} = \varphi{^A}_{B'}\textsf{h}^{B'B}$,
etc.\\

With these conventions,
if $\theta\in \hgm\otimes \Omega^1(\mathcal{Y})$ we write $\theta = \textbf{t}_A\theta^A$
and $[\theta\wedge \theta]_1 = \textbf{t}_A[\theta\wedge \theta]^A_1$
with $[\theta\wedge \theta]^A_1 := c^A_{BC}\theta^B\wedge \theta^C$ and hence
\[
d\theta^A + \frac{1}{2} [\theta\wedge \theta]^A_1 =d\theta^A + \frac{1}{2} c^A_{BC}\theta^B\wedge \theta^C
\]
If $\varphi \in so(\hgm,\textsf{h})\otimes \Omega^1(\mathcal{Y})\simeq \Lambda^2\hgm\otimes \Omega^1(\mathcal{Y})$ we write
$\varphi = \frac{1}{2}\textbf{t}_{AB}\varphi^{AB}$
and $[\varphi\wedge \varphi]_2 = \frac{1}{2}\textbf{t}_{AB}[\varphi\wedge \varphi]_2^{AB}$
with $[\varphi\wedge \varphi]_2^{AB}:= 
 2\textsf{h}_{A'B'} \varphi^{AA'}\wedge \varphi^{B'B}
 = 2\varphi{^A}_{A'}\wedge \varphi^{A'B}$ and hence
\[
d\varphi^{AB} + \frac{1}{2} [\varphi\wedge \varphi]_2^{AB} 
 = d\varphi^{AB} +\varphi{^A}_{A'}\wedge \varphi^{A'B}.
\]
Constraint (\ref{deuxiemecontrainte}) then
reads $\theta^a\wedge \theta^b\wedge \pi = 0$, $\forall a,b = 1,\cdots,n$.

\subsection{Useful relations}
Assume that the rank of $\theta\in \hgm\otimes \Omega^1(\mathcal{Y})$
is equal to $N$ everywhere and decompose $\theta = \textbf{t}_A\theta^A$.
Then $(\theta^1,\cdots,\theta^N)$ is a coframe on $\mathcal{Y}$. 
We denote
by $(\frac{\partial}{\partial \theta^1},\cdots,\frac{\partial}{\partial \theta^N})$
its dual frame.
We define recursively
\begin{equation}
 \theta^{(N-1)}_A:= \frac{\partial}{\partial \theta^A}\iN \theta^{(N)},\quad
 \theta^{(N-2)}_{AB}:=\frac{\partial}{\partial \theta^B}\iN \theta^{(N-1)}_A,\quad
 \theta^{(N-3)}_{ABC}:=\frac{\partial}{\partial \theta^C}\iN \theta^{(N-2)}_{AB}.
\end{equation}
Using the fact that $\theta^{(N)}
= \frac{1}{N!}\epsilon_{A_1\cdots A_N}\theta^{A_1}\wedge \cdots \wedge \theta^{A_N}$
one may prove that 
\begin{equation}\label{thetamoinsAalternatif}
 \theta^{(N-1)}_A = 
 \frac{1}{(N-1)!}\epsilon_{AA_2\cdots A_N}\theta^{A_2}\wedge \cdots
 \wedge \theta^{A_N},
\end{equation}
\begin{equation}\label{thetamoinsABalternatif}
 \theta^{(N-2)}_{AB} = 
 \frac{1}{(N-2)!}\epsilon_{ABA_3\cdots A_N}\theta^{A_3}\wedge \cdots
 \wedge \theta^{A_N},
\end{equation}
\begin{equation}\label{thetamoinsABCalternatif}
 \theta^{(N-3)}_{ABC} = 
 \frac{1}{(N-3)!}\epsilon_{ABCA_4\cdots A_N}\theta^{A_4}\wedge \cdots
 \wedge \theta^{A_N},\hbox{ etc.}
\end{equation}
Moreover we have the following
\begin{equation}\label{thetathetamoinsA}
 \theta^A\wedge \theta^{(N-1)}_{A'} = \delta^A_{A'}\theta^{(N)},
\end{equation}
\begin{equation}\label{thetathetamoinsAB}
 \theta^A\wedge \theta^{(N-2)}_{A'B'} = \delta^A_{B'}\theta^{(N-1)}_{A'}
 - \delta^A_{A'}\theta^{(N-1)}_{B'}
\end{equation}
and
\begin{equation}\label{thetathetamoinsABC}
 \theta^A\wedge \theta^{(N-3)}_{A'B'C'} = \delta^A_{C'}\theta^{(N-2)}_{A'B'}
 + \delta^A_{B'}\theta^{(N-2)}_{C'A'} + \delta^A_{A'}\theta^{(N-2)}_{B'C'}.
\end{equation}
Indeed (\ref{thetathetamoinsA}) can be proved by
developping the relation 
$0 = \frac{\partial}{\partial \theta^{A'}}\iN 0
= \frac{\partial}{\partial \theta^{A'}}\iN (\theta^A\wedge \theta^{(N)})$.
Computing the interior product by $\frac{\partial}{\partial \theta^{B'}}$
to both sides of (\ref{thetathetamoinsA}) leads to (\ref{thetathetamoinsAB}) and 
computing the interior product by $\frac{\partial}{\partial \theta^{C'}}$
to both sides of (\ref{thetathetamoinsAB}) leads to (\ref{thetathetamoinsABC}).

Lastly we have the following formulas
\begin{equation}\label{dthetamoinsA}
 d\theta^{(N-1)}_A = d\theta^B\wedge \theta^{(N-2)}_{AB},
\end{equation}
\begin{equation}\label{dthetamoinsAB}
 d\theta^{(N-2)}_{AB} = d\theta^C\wedge \theta^{(N-3)}_{ABC},
\end{equation}
which can be proved, e.g., by using (\ref{thetamoinsAalternatif}),
(\ref{thetamoinsABalternatif}) and (\ref{thetamoinsABCalternatif}).\\

As an application, assuming that the rank of $\theta\in \hgm\otimes \Omega^1(\mathcal{Y})$
is equal to $N$, we have
\[
 \star \theta^{(N-2)} = \frac{1}{2}\theta_{AB}^{(N-2)}\textbf{t}^{AB}.
\]
(Thus Condition (\ref{contraintebrute}) reads 
$\psi  = \frac{1}{2}\theta^{(N-2)}_{AB}\textbf{t}^{AB}$
or, equivalentely, $\theta^A\wedge \theta^B\wedge \psi = \textbf{t}^{AB}\theta^{(N)}$,
$\forall A,B = 1,\cdots,N$.)

\subsection{More precisions on the proof}
In the proof of the Theorem,
once we prove the existence of a fibration of 
$\mathcal{Y}$ over $\mathcal{X}$ and once
a local trivialization of this bundle has been chosen
(characterized by a projection map from $\mathcal{Y}$ to $\mathcal{X}$
and a map $g$ from an open subset of $\mathcal{Y}$ to $\mathfrak{G}_0$),
one can write that
$\theta^a = e^a$ and $\theta^\alpha = (gA g^{-1} + g^{-1}dg)^\alpha$, where $e^a$ and
$A = \textbf{t}_\alpha A^\alpha$ are pull-back
forms of 1-forms on $\mathcal{X}$.
Then a metric $\textbf{g}$ on $\mathcal{X}$ is defined by $\textbf{g} = (\theta^\mathfrak{s}) ^*\textsf{b} =
\textsf{b}_{ab}e^ae^b$ and $A$ is the expression of the 
connection $\nabla$ in the trivialization.
We then set $F:= dA +\frac{1}{2}[A\wedge A]$, the curvature 2-form of $A$.
The variation with respect to $\theta$ leads to the
equation
\[
 \frac{1}{2}\theta_{ABC}^{(-3)}\wedge \left(d\varphi^{AB}
 + \varphi{^A}_D\wedge \varphi^{DB}\right) = -d\pi_C -
 c^B_{CA}\theta^A\wedge \pi_B \hbox{ mod}[\theta_\gamma^{(N-1)}].
\]
One can recognize on the left hand side the Einstein tensor of $\mathbf{h}$ on $\mathcal{Y}$.
After a gauge transformation $e^\alpha = S^\alpha_\beta\theta^\beta$ and $\omega{^\alpha}_\beta
= S^\alpha_{\alpha'}\varphi{^{\alpha'}}_{\beta'}(S^{-1})^{\beta'}_\beta
- dS^\alpha_{\beta'}(S^{-1})^{\beta'}_\beta$, where $S = \hbox{Ad}_g$
(see Section \ref{transformationdejauge}), the previous equation translates as
\[
 \frac{1}{2}e_{ABC}^{(-3)}\wedge \left(d\omega^{AB}
 + \omega{^A}_D\wedge \omega^{DB}\right) = -dp_C \hbox{ mod}[e_\gamma^{(-1)}]
\]
The key observations are that the left hand side is constant on any fiber, whereas the restriction of the
right hand side to any fiber is an exact form. Both observations lead to the conclusion
that $\frac{1}{2}e_{ABC}^{(-3)}\wedge \left(d\omega^{AB}
 + \omega{^A}_C\wedge \omega^{CB}\right) = 0 \hbox{ mod}[e_\gamma^{(-1)}]$, i.e. the two
blocks $\hbox{Ein}(\textbf{h}){^a}_c$ and $\hbox{Ein}(\textbf{h}){^a}_\gamma$ of the Einstein
tensor of $\omega$ vanish.

The final equations, after a long computation (see Section \ref{calculfinal}) then read 

 \begin{equation}\label{eym0}
  \left\{
  \begin{array}{ccl}
     \hbox{Ein}(\textbf{g}){^a}_d + \frac{1}{2}
     \left(F{_\gamma}^{ac}F{\gamma}_{dc} - \frac{1}{4}F{_\gamma}^{bc}F{\gamma}_{bc})\delta^a_d\right)
     + \Lambda \delta^a_d
     & = & 0 \\
     \nabla_c F{_\gamma}^{ca} - c^\beta_{\alpha\gamma}A^\alpha_cF{_\beta}^{ca} & = & 0
    \end{array}
    \right. 
 \end{equation}
 where $\hbox{Ein}(\textbf{g}){^a}_d:= \hbox{Ric}(\textbf{g}){^a}_d - \frac{1}{2}\hbox{R}(\textbf{g})\delta^a_d$
 is the Einstein tensor of $\textbf{g}$, $F:= dA +\frac{1}{2}[A\wedge A]$ and
 $\Lambda:= - \frac{1}{8}c^\alpha_{\beta\gamma}c^\beta_{\alpha\epsilon} \textsf{h}^{\gamma\epsilon}
 = - \frac{1}{8}(K,\textsf{h}^*)$, where $K$ is the Killing form on $\mathfrak{g}$.


\section{The Euler--Lagrange equations}
In the following we assume that $(\theta,\varphi,\pi) \in \mathscr{C}$ 
is a critical point of $\mathscr{A}$ such that $\hbox{rank}\theta = N$
(Hypothesis (v)).
We denote by $\textbf{h} = \textsf{b}_{ab}\theta^a\theta^b
+ \textsf{k}_{\alpha\beta}\theta^\alpha\theta^\beta$ the induced metric on $\mathcal{Y}$
and we assume that $\textbf{h}$ is vertically complete (Hypothesis (vi)).

\subsection{Variations with respect to coefficients of $\pi$}
Since $\hbox{rank}\theta = N$, the family $(\theta^1,\cdots, \theta^{N})$
is a coframe on $\mathcal{Y}$, there exists unique coefficients $H^A_{BC}$ such that
$d\theta^A + \frac{1}{2} c^A_{BC}\theta^B\wedge \theta^C = \frac{1}{2}H^A_{BC}\theta^B\wedge \theta^C$
and $H^A_{BC} + H^A_{CB} = 0$.
We decompose $\pi = \pi_A\textbf{t}^A$ and each
$\pi_A$ as
$\pi_A = \frac{1}{2}\pi_A^{BC}\theta^{(N-2)}_{BC}$, where $\pi_A^{BC} + \pi_A^{CB} = 0$.
The constraint (\ref{deuxiemecontrainte}) then reads $\pi_A^{ab} = 0$ or
\begin{equation}\label{constraint1}
 \pi_A = \pi_A^{b\gamma}\theta^{(N-2)}_{b\gamma}
 + \frac{1}{2}\pi_A^{\beta\gamma}\theta^{(N-2)}_{\beta\gamma}
\end{equation}
A first order variation of $(\theta,\varphi,\pi)$ keeping $\theta$ and $\varphi$ constant
and respecting (\ref{constraint1}) thus induces a variation
of $\pi$ of the form
$\delta \pi_A = \chi_A^{b\gamma}\theta^{(N-2)}_{b\gamma}
 + \frac{1}{2}\chi_A^{\beta\gamma}\theta^{(N-2)}_{\beta\gamma}$.
The fact that the action $\mathscr{A}$ is stationary with respect to such variations
of $\pi$ thus reads
\[
 \int_\mathcal{Y} \delta \pi_A\wedge \left(d\theta^A + \frac{1}{2} [\theta\wedge \theta]^A \right)
 = \int_\mathcal{Y} \left(\chi_A^{b\gamma}H^A_{b\gamma}
 + \frac{1}{2}\chi_A^{\beta\gamma}H^A_{\beta\gamma}\right)\theta^{(N)} = 0,
 \quad \forall \chi_A^{b\gamma},\chi_A^{\beta\gamma}
\]
and lead to the Euler--Lagrange equations $H^A_{b\gamma} = H^A_{\beta\gamma} = 0$,
$\forall A,b,\beta,\gamma$. We thus deduce that
\begin{equation}\label{ELFrobenius}
\Theta^A:= 
 d\theta^A + \frac{1}{2} c^A_{BC}\theta^B\wedge \theta^C = \frac{1}{2}H^A_{bc}\theta^b\wedge \theta^c
\end{equation}
or equivalentely
\begin{equation}\label{ELFrobeniusdetaillee}
 \left\{\begin{array}{ccl}
         d\theta^a & = & \frac{1}{2}H^a_{bc}\theta^b\wedge \theta^c \\
         d\theta^\alpha + \frac{1}{2} c^\alpha_{\beta\gamma}\theta^\beta\wedge \theta^\gamma
         & = & \frac{1}{2}H^\alpha_{bc}\theta^b\wedge \theta^c
        \end{array}\right.
\end{equation}
\subsection{Variations with respect to $\varphi$}
Keeping $\theta$ and $\pi$ fixed we look at first order variations $\delta\varphi = \lambda$
of $\varphi$. This induces the condition that, for all $\lambda$,
\[
 \frac{1}{2}\int_\mathcal{Y}d\left(\lambda^{AB}\wedge \theta^{(N-2)}_{AB}\right)
 + \lambda^{AB}\wedge \left(d\theta^{(N-2)}_{AB}
 - \varphi{^{A'}}_A\wedge \theta^{(N-2)}_{A'B}
 - \varphi{^{B'}}_B\wedge \theta^{(N-2)}_{AB'} \right)
 = 0
\]
Assuming that $\lambda$ has compact support and using (\ref{thetathetamoinsABC})
and (\ref{dthetamoinsAB}) we deduce the relation
\[
 \left(d\theta^C + \varphi{^C}_{C'}\wedge \theta^{C'}\right)\wedge \theta^{(N-3)}_{ABC}
 = d\theta^{(N-2)}_{AB}
 - \varphi{^{A'}}_A\wedge \theta^{(N-2)}_{A'B}
 - \varphi{^{B'}}_B\wedge \theta^{(N-2)}_{AB'} 
 = 0
\]
which implies that the torsion 2-form $d\theta^A + \varphi{^A}_{A'}\wedge \theta^{A'}$ vanishes.
Hence the connection on $T\mathcal{Y}$ associated to $\varphi$ coincides with
the Levi-Civita connection of $(\mathcal{Y},\textbf{h})$, where
$\textbf{h} = \textsf{h}_{AB}\theta^A\theta^B$.
\subsection{Variations with respect to $\theta$}
We first observe that, through a variation $\delta\theta = \tau$ of $\theta$ keeping $\varphi$
and the coefficients $\pi_A^{b\gamma}$ and $\pi_A^{\beta\gamma}$ fixed, we have
\[
 \delta\left(\theta_{AB}^{(N-2)}\right) = \tau^C\wedge \theta^{(N-3)}_{ABC},
\]
plus the relation $\delta\Theta^A = d\tau^A+c^A_{BC}\tau^B\wedge \theta^C$
which implies 
\[
 \pi_A\wedge \delta\Theta^A 
 = d\left(\tau^A\wedge \pi_A\right) +\tau^A\wedge \left(d\pi_A + c^C_{AB}\theta^B\wedge \pi_C\right)
\]
and lastly
$\delta \pi_A = \pi_A^{b\gamma}\left(\tau^d\wedge \theta^{(N-3)}_{b\gamma d}
+ \tau^\delta \wedge \theta^{(N-3)}_{b\gamma\delta}\right)
+ \frac{1}{2}\pi_A^{\beta\gamma}\left(\tau^d\wedge \theta^{(N-3)}_{\beta\gamma d}
+ \tau^\delta \wedge \theta^{(N-3)}_{\beta\gamma\delta}\right)$
which, thanks to 
$\Theta^A \wedge \theta^{(N-3)}_{b\gamma\delta} =
\Theta^A \wedge \theta^{(N-3)}_{\beta\gamma d} =
\Theta^A \wedge \theta^{(N-3)}_{\beta\gamma\delta} = 0$
by (\ref{ELFrobenius}), leads to
\[
 \left(\delta \pi_A\right)\wedge \Theta^A 
 = - \pi_A^{b\gamma}H^A_{bd}\tau^d\wedge \theta^{(N-1)}_\gamma.
\]
In conclusion, by assuming that $\tau$ has compact support, we obtain
\[
 \int_\mathcal{Y}\tau^C\wedge\left( \frac{1}{2}\theta^{(N-3)}_{ABC}\wedge \Phi^{AB}
 - \pi_A^{b\gamma}H^A_{bC}\theta^{(N-1)}_\gamma
 + d\pi_C - c^B_{AC}\theta^A\wedge \pi_B\right) =  0
\]
where we set $\Phi:= d\varphi+\frac{1}{2}[\varphi\wedge \varphi]$.
Hence we deduce the Euler--Lagrange equation
\begin{equation}\label{ELEinsteinbrut}
 \frac{1}{2}\theta^{(N-3)}_{ABC}\wedge \Phi^{AB}
 + d\pi_C - c^B_{AC}\theta^A\wedge \pi_B
 = 0 \hbox{ mod}[\theta^{(N-1)}_\mathfrak{g}]
\end{equation}
where, for any 3-form $\lambda$, $\lambda = 0 \hbox{ mod}[\theta^{(N-1)}_\mathfrak{g}]$
means that there exists coefficients $\lambda^\alpha$ such that
$\lambda = \lambda^\alpha\theta^{(N-1)}_\alpha$.

\section{The fibration}

From the first equation in (\ref{ELFrobeniusdetaillee})
we deduce that $d\theta^a = 0\hbox{ mod}[\theta^b]$, $\forall a = 1,\cdots ,n$.
Since the rank of $(\theta^1,\cdots,\theta^n)$ is equal to $n$ everywhere,
Frobenius' theorem implies that, for any point $\textsf{y}\in \mathcal{Y}$, there
exists a neighbourhood of $\textsf{y}$ in which there exists a unique submanifold $\textsf{f}$
of dimension $r$ crossing $\textsf{y}$ such that $\theta^a|_\textsf{f} = 0$, $\forall a = 1,\cdots ,n$.
Hence $\mathcal{Y}$ is foliated by integral leaves of dimension $r$.

Consider on the product manifold $\mathcal{Y}\times \mathfrak{G}$ the $\mathfrak{g}$-valued 1-form 
$\tau := h^{-1}dh-\theta^\mathfrak{g}$, where $(\textsf{y},h)$ denotes
a point in $\mathcal{Y}\times \mathfrak{G}$ and 
where $\theta^\mathfrak{g}:= \textbf{t}_\alpha\theta^\alpha$.
It satisfies the identity $d\tau + d\theta^\mathfrak{g} +\frac{1}{2}[\theta^\mathfrak{g}\wedge\theta^\mathfrak{g}]
+ [\theta^\mathfrak{g}\wedge \tau] + \frac{1}{2}[\tau\wedge \tau] = 0$.
However the second equation in (\ref{ELFrobeniusdetaillee}) implies that,
for any integral leaf $\textsf{f}$,
$d\theta^\mathfrak{g} + \frac{1}{2}[\theta^\mathfrak{g}\wedge \theta^\mathfrak{g}]|_\textsf{f} = 0$
and thus
$d(\tau|_{\textsf{f}\times \mathfrak{G}}) = 0\hbox{ mod}[\tau]$.
Hence, again by Frobenius' theorem, 
for any $(\textsf{y}_0,g_0)\in \textsf{f}\times \mathfrak{G}$,
there exist a unique $r$-dimensional submanifold $\Gamma\subset \textsf{f}\times \mathfrak{G}$ which is
a solution of $\tau|_\Gamma=0$  and which contains $(\textsf{y}_0,g_0)$. This implies
the existence of a unique map $g$ (the graph of which is $\Gamma$)
from a neighbourhood of $\textsf{y}_0$ in $\textsf{f}$ to $\mathfrak{G}$
such that $g(\textsf{y}_0) = g_0$ and $dg-g\theta^\mathfrak{g}|_\textsf{f}= 0$.
Moreover $g$ is clearly invertible.

Consider any smooth path $\gamma:[0,1]\longrightarrow \mathfrak{G}$ such that $\gamma(0)=1_\mathfrak{G}$
and a point $\textsf{y}_0\in \mathcal{Y}$.
By Hypothesis (vi)
we can associate to it a unique path $u:[0,1]\longrightarrow \textsf{f}$ such that $u(0) = \textsf{y}_0$
and $(u,\gamma)^*\tau = 0$.
The image of $(u,\gamma)$ is contained in some integral submanifold $\Gamma$ 
which coincides locally with the graph of
an invertible map $g$ as previously. Thus
to any path homotopic to $\gamma$ in $\mathfrak{G}$ with fixed extremities it corresponds
a path homotopic to $u$ in $\textsf{f}$ with fixed extremities. Since $\mathfrak{G}$ is simply connected
we can thus define a unique map
$T:\mathfrak{G}\longrightarrow \textsf{f}$ such that $T(1_\mathfrak{G}) = \textsf{y}_0$
and $(T\times Id)^*\tau = 0$. Hence $\mathfrak{G}$ is a universal cover of $\textsf{f}$
and, in particular, since $\mathfrak{G}$ is compact
$\textsf{f}$ is \emph{compact}.

To any \emph{fixed} $x=(x^1,\cdots,x^n)\in \mathbb{R}^n$ we associated the vector field $X$ on $\mathcal{Y}$
defined by $X = x^a\frac{\partial}{\partial\theta^a}$. Let $\textsf{f}_0$
be some integral leaf. Let us assume that $x$ is in the unit ball $B^n$ of $\mathbb{R}^n$.
Since $\textsf{f}_0$ is compact there exists a neighbourhood $\mathcal{T}$
of $\textsf{f}_0$ in $\mathcal{Y}$ and some $\varepsilon>0$ such that
the flow map $(t,\textsf{y})\longmapsto e^{tX}(\textsf{y})$
is defined on $[-\varepsilon,\varepsilon]\times \mathcal{T}$.
We observe that, due to (\ref{ELFrobeniusdetaillee}), $L_X\theta^a = H^a_{bc}x^b\theta^c$, $\forall a$. 
Hence there exists functions $M^a_{bc}$ on $\mathcal{Y}$ (depending on $x$) such that
$\left(e^{tX}\right)^*\theta^a = M^a_c\theta^c$, $\forall a$.
For any leaf $\textsf{f}\subset \mathcal{T}$, let $\iota:\textsf{f}\longrightarrow \mathcal{Y}$ its
embedding map and $\iota_t:= e^{tX}\circ \iota$. Note that the image of $\iota_t$ is
$e^{tX}(\textsf{f})$. We have then
\[
 \iota_t^*\theta^a = \left(e^{tX}\circ \iota\right)^*\theta^a
 = \iota^*\left(e^{tX}\right)^*\theta^a = \iota^*\left(M^a_c\theta^c\right),
 \quad \forall a.
\]
Thus the 1-form $\textbf{t}_a\theta^a$ vanishes on $e^{tX}(\textsf{f})$ iff
it vanishes on $\textsf{f}$, i.e. $\textsf{f}$ is an integral leaf iff
$e^{tX}(\textsf{f})$ is also an integral leaf. As a consequence the map
$B^n\times \textsf{f}_0\ni (x,\textsf{y})\longmapsto
e^{\varepsilon x^a\frac{\partial}{\partial\theta^a}}(\textsf{y})$ is
a local diffeomorphism onto a neighbourhood of $\textsf{f}_0$, which provides
us with a local trivialization of the set of leaves. Hence
the set $\mathcal{X}$ of integral leaves has the structure of an $n$-dimensional manifold
and the quotient map $P:\mathcal{Y}\longrightarrow \mathcal{X}$ is a bundle fibration.

Set $e^a:= \theta^a$, for $1\leq a\leq n$.
From $\frac{\partial}{\partial \theta^\beta}\iN e^a = \frac{\partial}{\partial \theta^\beta}\iN de^a = 0$
we deduce that there exists a coframe $\left(\underline{e}^a\right)_{1\leq a\leq n}$ on $\mathcal{X}$
such that $e^a = P^*\underline{e}^a$, $\forall a$. Thus we can equipp $\mathcal{X}$ with the
pseudo Riemannian metric $\underline{\textbf{g}}:= \textsf{b}_{ab}\underline{e}^a\underline{e}^b$.

In the following we choose an $n$-dimensional submanifold $\Sigma \subset \mathcal{Y}$
transverse to the fibration. Without loss of generality
(replacing $\mathcal{Y}$ by an open subset of $\mathcal{Y}$ if necessary)
we can assume that $\Sigma$ intersects all fibers of $P$ and 
we define the map $g:\mathcal{Y}\longrightarrow \mathfrak{G}$ 
which is constant equal to $1_\mathfrak{G}$ on $\Sigma$ and
such that $dg-g\theta^\mathfrak{g}|_\textsf{f}=0$ for any
integral leaf $\textsf{f}$.
We then define $A:= g\theta^\mathfrak{g}g^{-1} - dg\cdot g^{-1}$.
The relation $dg-g\theta^\mathfrak{g}|_\textsf{f}=0$ then translates as
$A|_\textsf{f} = 0$ and hence we have the decomposition
$A = A_a\theta^a$.
Moreover since
\begin{equation}\label{thetaenfonctiondeA}
\theta^\mathfrak{g} = g^{-1}Ag + g^{-1}dg, 
\end{equation}
we have
$d\theta^\mathfrak{g} +\frac{1}{2}[\theta\wedge\theta]^\mathfrak{g} = g^{-1}(dA+\frac{1}{2}[A\wedge A])g
= g^{-1}Fg$, where $F:= dA + \frac{1}{2}[A\wedge A]$.
By using (\ref{ELFrobeniusdetaillee}) we deduce that $\frac{\partial}{\partial \theta^\alpha}\iN
dA = 0$, $\forall \alpha=n+1,\cdots,N$, i.e. the coefficients $A_a$ are constants on the fibers
$\textsf{f}$. Moreover we have
\begin{equation}\label{decompositiondeF}
 F^\alpha = \frac{1}{2}F^\alpha_{bc}e^b\wedge e^c,
\end{equation}
where the coefficients $F_{bc} = gH_{bc}^\mathfrak{g}g^{-1}$ are constant on the fibers.

\section{Trivialization of the bundle}\label{transformationdejauge}
Using the map $g:\mathcal{Y}\longrightarrow \mathfrak{G}$ defined previously
we define the map $S$ from $\mathcal{Y}$ to $\hbox{End}(\hgm)$
which, to any $\textsf{y}\in \mathcal{Y}$, associates $\hbox{Ad}_{g(\textsf{y})}$.
In other words, $\forall (v,\xi)\in \mathfrak{s}\times \mathfrak{g}$,
$S(v+\xi):= g(v+\xi)g^{-1} = v + g\xi g^{-1}$.
We remark that $S$ takes values in $SO(\hgm,\textsf{h})$ because of
Hypothesis (ii).
Let $\left(S^A_B\right)_{1\leq A,B\leq N}$ be the matrix of $S$ in the basis 
$\left(\textbf{t}_A\right)_{1\leq A\leq N}$, i.e. such that
$S(\textbf{t}_A) = \textbf{t}_BS^B_A$.
We define a new coframe
$\left(e^A\right)_{1\leq A\leq N}$ by $e^A = S^A_B\theta^B$.
Equivalentely 
\[
 \left\{\begin{array}{cccl}
         e^a & := & \theta^a & \forall a = 1,\cdots,n \\
         e^\alpha & := & S^\alpha_\beta \theta^\beta & \forall \alpha = n+1,\cdots,N
        \end{array}\right.
\]
Then $e^\alpha = (g\theta^\mathfrak{g}g^{-1})^\alpha$ and
(\ref{thetaenfonctiondeA}) imply
\begin{equation}\label{formulepoure}
 e^\alpha = A^\alpha + (dg\ g^{-1})^\alpha,\quad \forall \alpha = n+1,\cdots,N.
\end{equation}
We deduce that
\[
\begin{array}{ccl}
 de^\alpha - \frac{1}{2} [e\wedge e]^\alpha + [A\wedge e]^\alpha
 & = & de^\alpha - \frac{1}{2} [e\wedge e]^\alpha + [(e-dg\,g^{-1})\wedge e]^\alpha \\
 & = & de^\alpha + \frac{1}{2} [e\wedge e]^\alpha - [dg\,g^{-1}\wedge e]^\alpha \\
 & = & \left(gd\theta g^{-1} + [dgg^{-1}\wedge e]\right)^\alpha
 + \frac{1}{2}[e \wedge e]^\alpha - [dgg^{-1}\wedge e]^\alpha \\
 & = & \left(g\left(d\theta + \frac{1}{2}[\theta\wedge \theta]\right)g^{-1}\right)^\alpha
\end{array}
\]
from which we get the useful identity
\begin{equation}\label{deextra}
 de^\alpha - \frac{1}{2} [e\wedge e]^\alpha + [A\wedge e]^\alpha
 = F^\alpha := \frac{1}{2}F_{bc}^\alpha e^b\wedge e^c.
\end{equation}

Let us translate the left hand side of (\ref{ELEinsteinbrut}) in the new coframe.
First we define $e^{(N)}:= e^1\wedge \cdots \wedge e^{N}$ and note that $e^{(N)} = \theta^{(N)}$.
Moreover defining $e^{(N-1)}_A:= \frac{\partial}{\partial e^A}\iN e^{(N)}$,
$e^{(N-2)}_{AB}:= \frac{\partial}{\partial e^A}\wedge\frac{\partial}{\partial e^B}\iN e^{(N)}$,
we observe that, since $\frac{\partial}{\partial \theta^A} = \frac{\partial}{\partial e^B}S^B_A$,
we have $\theta^{(N-1)}_A = e^{(N-1)}_{A'}S^{A'}_A$,
$\theta^{(N-2)}_{AB} = e^{(N-2)}_{A'B'}S^{A'}_AS^{B'}_B$ and
$\theta^{(N-3)}_{ABC} = e^{(N-3)}_{A'B'C'}S^{A'}_AS^{B'}_BS^{C'}_C$.

Second let $\omega$ be the $so(\hgm,\textsf{h})$-valued connection 1-form 
in the coframe $(e^A)_{1\leq A\leq N}$, which is equal to 
$\omega:= S\varphi S^{-1} - dS\, S^{-1}$. 
Let $\Omega:= d\omega + \frac{1}{2}[\omega\wedge\omega] = S\Phi S^{-1}$,
where $\Phi = d\varphi + \frac{1}{2}[\varphi\wedge\varphi]$.
Then $\Phi^{AB} = (S^{-1})^A_{A'}(S^{-1})^B_{B'}\Omega^{A'B'}$.

We deduce that $\theta^{(N-3)}_{ABC}\wedge \Phi^{AB} = e^{(N-3)}_{ABC'}\wedge \Omega^{AB}S^{C'}_C$.
Hence (\ref{ELEinsteinbrut}) is equivalent to
\begin{equation}\label{ELEinsteinbrut2}
 \frac{1}{2} e^{(N-3)}_{ABC}\wedge \Omega^{AB} + \left(d\pi_{C'} - c^B_{AC'}\theta^A\wedge \pi_B\right)(S^{-1})^{C'}_C
 = 0\hbox{ mod}[e^{(N-1)}_\mathfrak{g}],
\end{equation}
where, for any $(N-1)$-form $\lambda$, we write:
\[
 \lambda = 0\hbox{ mod}[e^{(N-1)}_\mathfrak{g}]
\]
iff there exists forms $\lambda^\alpha$ such that $\lambda = \lambda^\alpha e^{(N-1)}_\alpha$.

In the following we use extensively Relations (\ref{thetathetamoinsA}),
(\ref{thetathetamoinsAB}), (\ref{thetathetamoinsABC}), (\ref{dthetamoinsA})
and (\ref{dthetamoinsAB}).
\begin{lemma}
 We have
 \begin{equation}\label{formuledulemme}
  \left(d\pi_{C'} - c^B_{AC'}\theta^A\wedge \pi_B\right)(S^{-1})^{C'}_C
  = d\left(\pi_{C'}(S^{-1})^{C'}_C\right) \hbox{ mod}[e^{(N-1)}_\mathfrak{g}].
 \end{equation}
\end{lemma}
\emph{Proof} --- From the definition of $S$ we deduce that, $\forall \xi\in \hgm$,
\[
 d(S^{-1}(\xi)) = [g^{-1}\xi g,g^{-1}dg] = g^{-1}[\xi,\textbf{t}_\alpha(e^\alpha - A^\alpha)]g
 = - S^{-1}([e-A,\xi]),
\]
where, in the last equality we used the fact that $[\xi,\textbf{t}_ae^a]= 0$ because of Hypothesis (i).
Thus we can write
$d(S^{-1})^{C'}_C = - (S^{-1})^{C'}_Ac^A_{BC}(e^B-A^B)$.
Hence
\[
\begin{array}{ccl}
 d\left((S^{-1})^{C'}_C\pi_{C'}\right) & = &
 - (S^{-1})^{C'}_Ac^A_{BC}(e^B-A^B)\wedge \pi_{C'} + (S^{-1})^{C'}_C\left(d\pi_{C'}\right) \\
  & = & - (S^{-1})^{C'}_Ac^A_{BC}S^B_{B'}\theta^{B'}\wedge \pi_{C'}
  + (S^{-1})^{C'}_Ac^A_{BC}A^B\wedge \pi_{C'}+ (S^{-1})^{C'}_C\left(d\pi_{C'}\right)
\end{array}
\]
But because of  $[\hbox{Ad}_g(\xi),\hbox{Ad}_g(\eta)] = \hbox{Ad}_g([\xi,\eta])$,
$\forall \xi,\eta\in \mathfrak{g}$, which is equivalent to $c^A_{B'C'}S^{B'}_BS^{C'}_C
= S^A_{A'}c^{A'}_{BC}$, we have $(S^{-1})^{C'}_Ac^A_{BC}S^B_{B'} = c^{C'}_{B'C''}(S^{-1})^{C''}_C$.
Thus for the first term on the r.h.s.,
\[
 (S^{-1})^{C'}_Ac^A_{BC}S^B_{B'}\theta^{B'}\wedge \pi_{C'}
= c^{C'}_{B'C''}(S^{-1})^{C''}_C\theta^{B'}\wedge \pi_{C'}
= c^{B}_{AC'}\theta^{A}\wedge \pi_{B} (S^{-1})^{C'}_C
\]
and hence
\[
d\left((S^{-1})^{C'}_C\pi_{C'}\right) = \left(d\pi_{C'} - c^B_{AC'}\theta^A\wedge \pi_B\right)(S^{-1})^{C'}_C
+ (S^{-1})^{C'}_Ac^A_{BC}A^B\wedge \pi_{C'}
\]
However it follows from (\ref{constraint1}) that 
$\pi_{C'} = S^\gamma_{\gamma'}\pi^{b\gamma'}_{C'}e^{(N-2)}_{b\gamma}
+ \frac{1}{2}S^\beta_{\beta'}S^\gamma_{\gamma'}\pi^{\beta'\gamma'}_{C'}
e^{(N-2)}_{\beta\gamma}$ and, since $A^B = A^B_ce^c$, we
get $(S^{-1})^{C'}_Ac^A_{BC}A^B\wedge \pi_{C'}
= -(S^{-1})^{C'}_Ac^A_{BC}A^B_b S^\gamma_{\gamma'}\pi_{C'}^{b\gamma'}e^{(N-1)}_\gamma
= 0\hbox{ mod}[e^{(N-1)}_\mathfrak{g}]$.
Hence (\ref{formuledulemme}) follows. \hfill $\square$\\

\noindent
Thus if we define $p_C:= \pi_{C'}(S^{-1})^{C'}_C$ we deduce from (\ref{formuledulemme})
that (\ref{ELEinsteinbrut2}) is equivalent to
\begin{equation}\label{ELEinsteinbrut3}
 \frac{1}{2} e^{(N-3)}_{ABC}\wedge\Omega^{AB} + dp_C = 0\hbox{ mod}[e^{(N-1)}_\mathfrak{g}].
\end{equation}

We need to compute $dp_C$. For that purpose we use the 
\emph{a priori} decomposition $p_C = p^{b\gamma}_{C}e^{(N-2)}_{b\gamma}
+ \frac{1}{2}p^{\beta\gamma}_{C}e^{(N-2)}_{\beta\gamma}$. We first compute
using (\ref{thetathetamoinsAB}), (\ref{thetathetamoinsABC}),
(\ref{ELFrobeniusdetaillee}) and (\ref{deextra})
\[
 \begin{array}{ccl}
  de^{(N-2)}_{b\gamma} & = & de^a\wedge e_{b\gamma a}^{(N-3)} + de^\alpha\wedge e_{b\gamma \alpha}^{(N-3)} \\
  & = & H^a_{ab}e_\gamma^{(N-1)} + c^\alpha_{\gamma\alpha}e_b^{(N-1)}
  - c^\alpha_{\beta\gamma} (A^\beta)_be_\gamma^{(N-1)} \\
  & = & \left(H^a_{ab} - c^\alpha_{\beta\gamma} (A^\beta)_b\right)e_\gamma^{(N-1)}
  = 0\hbox{ mod}[e^{(N-1)}_\mathfrak{g}],
 \end{array}
\]
where we have used the fact that, since $\mathfrak{G}$ is compact, its Lie algebra $\mathfrak{g}$
is \emph{unimodular}, which reads $c^\alpha_{\gamma\alpha} = 0$. Similarly
\[
 \begin{array}{ccl}
  de^{(N-2)}_{\beta\gamma} & = & de^a\wedge e_{\beta\gamma a}^{(N-3)}
  + de^\alpha\wedge e_{\beta\gamma \alpha}^{(N-3)} \\
  & = & 0 + c^\alpha_{\gamma\alpha}e_\beta^{(N-1)} + c^\alpha_{\alpha\beta}e_\gamma^{(N-1)}
  + c^\alpha_{\beta\gamma}e_\alpha^{(N-1)}  \\
  & = & 0\hbox{ mod}[e^{(N-1)}_\mathfrak{g}],
 \end{array}
\]
Thus by
writing $dp^{b\gamma}_{C} = p^{b\gamma}_{C,c}e^c + p^{b\gamma}_{C,\gamma}e^\gamma$
and $dp^{\beta\gamma}_{C} = p^{\beta\gamma}_{C,c}e^c + p^{\beta\gamma}_{C,\delta}e^\delta$,
we get
\[
 dp_C = p^{b\gamma}_{C,\gamma}e^{(N-1)}_b - p^{b\gamma}_{C,b}e^{(N-1)}_{\gamma}
+ p^{\beta\gamma}_{C,\gamma}e^{(N-1)}_{\beta} \hbox{ mod}[e^{(N-1)}_\mathfrak{g}] 
 = p^{b\gamma}_{C,\gamma}e^{(N-1)}_b
\hbox{ mod}[e^{(N-1)}_\mathfrak{g}]
\]

Lastly by decomposing $\Omega^{AB} = \frac{1}{2}\Omega{^{AB}}_{CD}e^C\wedge e^D$,
we find that
\[
 \frac{1}{2} e^{(N-3)}_{ABC}\wedge \Omega^{AB} =
- \hbox{Ein}(\omega){^A}_Ce^{(N-1)}_A,
\]
where
$\hbox{Ein}(\omega){^A}_C:= \hbox{Ric}(\omega){^A}_C - \frac{1}{2}\hbox{R}(\omega)\delta{^A}_C$,
$\hbox{Ric}(\omega){^A}_C:= \Omega{^{AB}}_{CB}$  and
$\hbox{R}(\omega):= \hbox{Ric}(\omega){^A}_A$. 
Obviously $\hbox{Ric}(\omega){^A}_C$ is the Ricci curvature,
$\hbox{R}(\omega)$ the scalar curvature and
$\hbox{Ein}(\omega){^A}_C$ the Einstein tensor of $\textsf{h}$
in the coframe $\left(e^A\right)_{1\leq A\leq N}$.
Hence  we find that (\ref{ELEinsteinbrut3}) is equivalent to
$\left(\hbox{Ein}(\omega){^a}_C - p^{a\gamma}_{C,\gamma}\right) e^{(N-1)}_a = 0
\hbox{ mod}[e^{(N-1)}_\mathfrak{g}]$,
or
\begin{equation}\label{ELEinsteinrefined}
 \hbox{Ein}(\omega){^a}_C = p^{a\gamma}_{C,\gamma} ,\quad \forall a=1,\cdots,n,\forall C = 1,\cdots, N.
\end{equation}
We will come back to this equation later on.

\section{Computation of the connection and the curvature forms}\label{calculfinal}
We need to compute the connection 1-form $\omega$ and its curvature 2-form.
As a preliminary we first set $\underline{\gamma}{^a}_c$ to be the connection 1-form
on $(\mathcal{X},\underline{\textbf{g}})$ in the coframe $\underline{e}^a$, i.e. which satisfies 
$\underline{\gamma}^{ac} + \underline{\gamma}^{ca} =0$ and
$d\underline{e}^a + \underline{\gamma}{^a}_c\wedge \underline{e}^b = 0$.
Then we set $\gamma{^a}_c := P^*\underline{\gamma}{^a}_c$ which satisfies similar
relations, which, together with (\ref{deextra}), leads to
\begin{equation}\label{formuledetrans}
\left\{\begin{array}{ccc}
de^a + \gamma{^a}_c\wedge e^c & = & 0 \\
 de^\alpha - \frac{1}{2}F^\alpha_{bc}e^b\wedge e^c
 - \frac{1}{2} c^\alpha_{\beta\gamma}(e^\beta - 2A^\beta)\wedge e^\gamma
 & = & 0
       \end{array} \right.
\end{equation}

Now the connexion 1-form $\omega$ is uniquely characterized by the condition $\omega^{AB}+\omega^{BA} = 0$
(preservation of the metric) and $de^A + \omega{^A}_C\wedge e^C = 0$ (the torsion vanishes),
which can be written
\begin{equation}\label{conditionssuromega}
 \left\{ \begin{array}{ccl}
          de^a + \omega{^a}_c \wedge e^c + \omega{^a}_\gamma \wedge e^\gamma & = & 0 \\
          de^\alpha + \omega{^\alpha}_c \wedge e^c + \omega{^\alpha}_\gamma \wedge e^\gamma & = & 0
         \end{array}\right.
\end{equation}
Comparing with (\ref{formuledetrans}) we are tempted to assume that 
$\omega{^\alpha}_\gamma = - \frac{1}{2} c^\alpha_{\beta\gamma}(e^\beta - 2A^\beta)$,
which fulfills the condition $\omega^{\alpha\beta} + \omega^{\beta\alpha} = 0$,
since
$c^\alpha_{\beta\gamma'}\textsf{k}^{\gamma'\gamma} + c^\gamma_{\beta\alpha'}\textsf{k}^{\alpha'\alpha} = 0$
because the metric $\textsf{k}$ is preserved by the adjoint action of $\mathfrak{g}$.
We also guess that $\omega{^\alpha}_c = - \frac{1}{2}F^\alpha_{bc}e^b$, which forces automatically
$\omega{^a}_\gamma = \frac{1}{2}\textsf{k}_{\gamma\gamma'}F^{\gamma'}_{bc'}\textsf{g}^{c'c}e^b$,
in order to satisfy $\omega^{\alpha b} + \omega^{b\alpha} = 0$.
Then in order to fulfill the first relation of (\ref{conditionssuromega}), one needs to
assume that $\omega{^a}_c = \gamma{^a}_c
- \frac{1}{2}\textsf{k}_{\gamma\gamma'}F^{\gamma'}_{a'c}\textsf{b}^{a'a}e^\gamma$.
We then check that $\omega^{ac} = \gamma^{ac} - 
\frac{1}{2}\textsf{k}_{\gamma\gamma'}F^{\gamma'}_{a'c'}\textsf{b}^{a'a}\textsf{b}^{c'c}e^\gamma$
is skew symmetric in $(a,c)$. Thus we see that the forms $\omega^{AC}$ defined by:
\[
 \left(\begin{array}{cc}
        \omega{^a}_c & \omega{^a}_\gamma \\
        \omega{^\alpha}_c & \omega{^\alpha}_\gamma
       \end{array}\right)
 = \left(\begin{array}{cc}
        \gamma{^a}_c
- \frac{1}{2}\textsf{k}_{\gamma\gamma'}F^{\gamma'}_{a'c}\textsf{b}^{a'a}e^\gamma
& \frac{1}{2}\textsf{k}_{\gamma\gamma'}F^{\gamma'}_{ba'}\textsf{b}^{a'a}e^b \\
        - \frac{1}{2}F^\alpha_{bc}e^b & - \frac{1}{2} c^\alpha_{\beta\gamma}(e^\beta - 2A^\beta)
       \end{array}\right)
\]
satisfy (\ref{conditionssuromega}) and $\omega^{AC} + \omega^{CA} = 0$. 
Hence this is the Levi-Civita connection 1-form of $(\mathcal{Y},\textbf{h})$.
In the following it will convenient to set 
$F{^\gamma}_{bc}:= F^\gamma_{bc}$, $F{_\gamma}{^a}_c:= \textsf{k}_{\gamma\gamma'}F{^{\gamma'}}_{a'c}\textsf{b}^{a'a}$
and $F{_{\gamma b}}^c:= \textsf{k}_{\gamma\gamma'}F{^{\gamma'}}_{bc'}\textsf{b}^{c'c}$.
Then
\[
 \left(\begin{array}{cc}
        \omega{^a}_c & \omega{^a}_\gamma \\
        \omega{^\alpha}_c & \omega{^\alpha}_\gamma
       \end{array}\right)
 = \left(\begin{array}{cc}
        \gamma{^a}_c
- \frac{1}{2}F{_{\gamma}}{^a}_{c}e^\gamma
& \frac{1}{2}F{_{\gamma b}}^{a}e^b \\
        - \frac{1}{2}F{^\alpha}_{bc}e^b & - \frac{1}{2} c^\alpha_{\beta\gamma}(e^\beta - 2A^\beta)
       \end{array}\right)
\]
We can thus compute the curvature 2-form $\Omega{^A}_C = d\omega{^A}_C +  \omega{^A}_B\wedge \omega{^B}_C$.

\[
 \Omega{^a}_c = d\left(\gamma{^a}_c - \frac{1}{2}F{_\gamma}{^a}_ce^\gamma \right)
 + \left(\gamma{^a}_b - \frac{1}{2}F{_{\beta'}}{^a}_be^{\beta'} \right)\wedge
 \left(\gamma{^b}_c - \frac{1}{2}F{_{\gamma'}}{^b}_ce^{\gamma'} \right)
 - \frac{1}{4}F{_{\beta b'}}^aF{^\beta}_{c'c} e^{b'}\wedge e^{c'}
\]
\[
 \Omega{^a}_\gamma = d\left(F{_{\gamma b}}^a e^b\right) 
 + \frac{1}{2}\left(\gamma{^a}_b - \frac{1}{2}F{_{\beta'}}{^a}_be^{\beta'} \right)\wedge
 \left(F{_{\gamma b'}}^a e^{b'}\right) 
 + \frac{1}{4}F{_{\beta b'}}^a c^\beta_{\beta'\gamma} e^{b'}\wedge (2A^{\beta'} - e^{\beta'})
\]
\[
 \Omega{^\alpha}_c = - d\left(F{^{\alpha}}_{bc} e^b\right) 
 - \frac{1}{2}\left(F{^{\alpha}}_{b'b} e^{b'}\right) \wedge
  \left(\gamma{^b}_c - \frac{1}{2}F{_{\gamma'}}{^b}_c e^{\gamma'} \right)
 - \frac{1}{4}F{^{\beta}}_{bc} c^\alpha_{\beta'\beta} (2A^{\beta'} - e^{\beta'})\wedge e^{b}
\]
\[
 \Omega{^\alpha}_\gamma = \frac{1}{2} d\left(c^\alpha_{\beta\gamma} (2A^{\beta} - e^{\beta})\right) 
 - \frac{1}{4}\left(F{^{\alpha}}_{b'b} e^{b'}\right) \wedge \left(F{_{\gamma c'}}^{b} e^{c'}\right)
 + \frac{1}{4} c^\alpha_{\beta'\beta}c^\beta_{\gamma'\gamma} (2A^{\beta'} - e^{\beta'})\wedge (2A^{\gamma'} - e^{\gamma'}).
\]
Lastly we obtain the components of the Ricci tensor $\hbox{Ric}(\omega)$ through a lengthy computation.
\begin{equation}\label{Ricciad}
 \hbox{Ric}(\omega){^a}_d = \hbox{Ric}(\gamma){^a}_d - \frac{1}{2}F{_\beta}^{ac}F{^\beta}_{dc}
\end{equation}
where $\hbox{Ric}(\gamma){^a}_d:= \left(d\gamma{^a}_c + \gamma{^a}_b\wedge \gamma{^b}_c\right)_{de}\textsf{b}^{ce}$
is the Ricci curvature of $\gamma$, and using the decompositions $dF_{\delta}{^{ac}} = F_{\delta}{^{ac}}_{,c}e^c$
and $\gamma{^a}_b = (\gamma{^a}_b)_ce^c$,
\begin{equation}\label{Ricciadelta}
 \hbox{Ric}(\omega){^a}_\delta = \frac{1}{2}\left( F_{\delta}{^{ac}}_{,c}
 + (\gamma{^a}_b)_c F{_\delta}^{bc} + (\gamma{^c}_b)_c F{_\delta}^{ab}
 - c^\gamma_{\alpha\delta}A^\alpha_c F{_\gamma}^{ac}\right)
\end{equation}
\begin{equation}\label{Riccialphad}
 \hbox{Ric}(\omega){^\alpha}_\delta = \frac{1}{4} F{_\delta}{^{bc}}F{^\alpha}{_{bc}}
 - \frac{1}{4} c^\alpha_{\beta\gamma} c^\beta_{\delta\epsilon}\textsf{k}^{\gamma\epsilon}
\end{equation}
We deduce the scalar curvature $\hbox{R}(\omega)$ of $\omega$ in function of the scalar curvature
$\hbox{R}(\gamma):= \hbox{Ric}(\gamma){^a}_a$:
\begin{equation}\label{Rscalaire}
 \hbox{R}(\omega) = \hbox{R}(\gamma)
 - \frac{1}{4} F{_\alpha}^{ab}F{^\alpha}_{ab}
 - \frac{1}{4} c^\alpha_{\beta\gamma} c^\beta_{\alpha\delta}\textsf{k}^{\gamma\delta}
\end{equation}
Hence the Einstein tensor of $\omega$ is
\begin{equation}\label{Einsteinad}
 \hbox{Ein}(\omega){^a}_d = \hbox{Ein}(\gamma){^a}_d - \frac{1}{2}\left(F{_\beta}^{ac}F{^\beta}_{dc}
 - \frac{1}{4} F{_\alpha}^{bc}F{^\alpha}_{bc}\delta{^a}_d\right)
 + \frac{1}{8}c^\alpha_{\beta\gamma} c^\beta_{\alpha\delta}\textsf{k}^{\gamma\delta} \delta{^a}_d
\end{equation}
and $\hbox{Ein}(\omega){^a}_\delta = \hbox{Ric}(\omega){^a}_\delta$ is given by (\ref{Ricciadelta}).

An important observation is that \emph{the components of} $\hbox{Ein}(\omega){^a}_d$
\emph{and} $\hbox{Ein}(\omega){^a}_\delta$ \emph{are
constant on the fibers $\textsf{f}$}.

\section{The Einstein--Yang--Mills equations}

We conclude by exploiting the fact that the fibers $\textsf{f}$ are compact without boundary.
Let $\mu^{(r)}:= e^{n+1}\wedge \cdots e^{N}$ and set
$\mu^{(r-1)}_\alpha:= \frac{\partial}{\partial e^\alpha}\iN \mu^{(r)}$, $\forall \alpha$.
By integrating both sides of (\ref{ELEinsteinrefined}) on a fiber $\textsf{f}$ we obtain
\[
 \int_\textsf{f}\hbox{Ein}(\omega){^a}_C \mu^{(r)} =  \int_\textsf{f}p^{a\gamma}_{C,\gamma} \mu^{(r)}
 = \int_\textsf{f}d\left(p^{a\gamma}_C \mu^{(r-1)}_\gamma\right) = 0.
\]
But on the one hand, the components of $\hbox{Ein}(\omega){^a}_C$ are constant on the fiber $\textsf{f}$,
as seen in the previous section.  Hence

\begin{equation}\label{ELEinsteinevident}
 \hbox{Ein}(\omega){^a}_C = \frac{\int_\textsf{f}\hbox{Ein}(\omega){^a}_C \mu^{(r)}}{\int_\textsf{f} \mu^{(r)}}
 =0.
\end{equation}
Thus, using (\ref{Ricciadelta}) and (\ref{Einsteinad}) we deduce that $\gamma$ and $A$ are solutions
of the Einstein--Yang--Mills system
\[
 \left\{ \begin{array}{ccl}
          \hbox{Ein}(\gamma){^a}_d - \frac{1}{2}\left(F{_\beta}^{ac}F{^\beta}_{dc}
 - \frac{1}{4} F{_\alpha}^{bc}F{^\alpha}_{bc}\delta{^a}_d\right)
 + \frac{1}{8}c^\alpha_{\beta\gamma} c^\beta_{\alpha\delta}\textsf{k}^{\gamma\delta} \delta{^a}_d
 & = & 0\\
 F_{\delta}{^{ac}}_{,c}
 + (\gamma{^a}_b)_c F{_\delta}^{bc} + (\gamma{^c}_b)_c F{_\delta}^{ab}
 - c^\gamma_{\alpha\delta}A^\alpha_c F{_\gamma}^{ac} & = & 0
         \end{array}\right.
\]

\bibliographystyle{plain}
\bibliography{bibkaluza}

\begin{thebibliography}{10}

\bibitem{appelquist}
T.~Appelquist, A.~Chodos, and P.G.O. Freund, editors.
\newblock {\em Modern Kaluza-Klein theories}.
\newblock Addison-Wesley Pub. Co., 1987.

\bibitem{chofreund}
Y.~M. Cho and Peter G.~O. Freund.
\newblock Non-abelian gauge fields as {N}ambu-{G}oldstone fields.
\newblock {\em Physical Review D}, 12(6):1711--1720, sep 1975.

\bibitem{cremmer-scherk}
E.~Cremmer and J.~Scherk.
\newblock Spontaneous compactification of space in an
  {E}instein-{Y}ang-{M}ills-{H}iggs model.
\newblock {\em Nuclear Physics B}, 108(3):409--416, jun 1976.

\bibitem{dewitt}
B.S. DeWitt.
\newblock Problem 77.
\newblock In {\em In Dynamical Theory of Groups and Fields}. Les {H}ouches,
  1963.
\newblock Reprinted in \cite{appelquist}.

\bibitem{einsteinbergmann}
A.~Einstein and P.~Bergmann.
\newblock On a generalization of {K}aluza's theory of electricity.
\newblock {\em The Annals of Mathematics}, 39(3):683, jul 1938.
\newblock Reprinted in \cite{appelquist}.

\bibitem{ferraris}
M.~Ferraris, M.~Francaviglia, and C.~Reina.
\newblock Variational formulation of general relativity from 1915 to 1925 ?
  "{P}alatini's method" discovered by {E}instein in 1925.
\newblock {\em General Relativity and Gravitation}, 14(3):243--254, mar 1982.

\bibitem{helein14}
Fr{\'{e}}d{\'{e}}ric H{\'{e}}lein.
\newblock Multisymplectic formulation of {Y}ang-{M}ills equations and
  {E}hresmann connections.
\newblock {\em Advances in Theoretical and Mathematical Physics},
  19(4):805--835, 2015.
\newblock Available in \href{https://arxiv.org/abs/1406.3641}{arXiv:1406.3641}.

\bibitem{heleinvey15}
Fr{\'{e}}d{\'{e}}ric H{\'{e}}lein and Dimitri Vey.
\newblock Curved space-times by crystallization of liquid fiber bundles.
\newblock {\em Foundations of Physics}, 47(1):1--41, sep 2016.
\newblock Available in
  \href{https://arxiv.org/abs/1508.07765}{arXiv:1508.07765} and
  \href{https://hal.archives-ouvertes.fr/hal-01205784v2}{hal-01205784v2}.

\bibitem{jordan}
P.~Jordan.
\newblock Erweiterung der projektiven {R}elativit\"{a}tstheorie.
\newblock {\em Annalen der Physik}, 436(4-5):219--228, 1947.

\bibitem{kaluza}
Th. Kaluza.
\newblock On the unification problem in physics.
\newblock {\em Sitzungsberichte Pruss. Acad. Sci.}, page 966, 1921.
\newblock Reprinted in English in \cite{appelquist} and available in
  \href{https://arxiv.org/abs/1803.08616}{arXiv:1803.08616}.

\bibitem{kerner}
R.~Kerner.
\newblock Generalization of the {K}aluza-{K}lein theory for an arbitrary
  non-{A}belian gauge group.
\newblock {\em Ann. Inst. H. Poincar{\'e}}, 9(2):143, 1968.
\newblock Available in
  \href{http://www.numdam.org/item?id=AIHPA_1968__9_2_143_0}{www.numdam.org}.

\bibitem{klein}
Oskar Klein.
\newblock Quantentheorie und f{\"u}nfdimensionale {R}elativit{\"a}tstheorie.
\newblock {\em Zeitschrift f{\"u}r Physik}, 37(12):895--906, dec 1926.

\bibitem{mandel}
Heinrich Mandel.
\newblock Zur {H}erleitung der {F}eldgleichungen in der allgemeinen
  {R}elativit{\"a}tstheorie.
\newblock {\em Zeitschrift f{\"u}r Physik}, 39(2-3):136--145, feb 1926.

\bibitem{nordstrom}
G.~Nordstr{\"o}m.
\newblock {\"U}ber die {M}{\"o}glichkeit, das elektromagnetische {F}eld und
  dans {G}ravitationsfeld zu vereinigen.
\newblock {\em Phys. Zeitsch.}, 15(504), 1914.
\newblock Available in
  \href{http://publikationen.ub.uni-frankfurt.de/frontdoor/index/index/docId/17520}{publikationen.ub.uni-frankfurt.de}.

\bibitem{thiry}
Y.~Thiry.
\newblock Les {\'e}quations de la th{\'e}orie unitaire de {K}aluza.
\newblock {\em Comptes Rendus Acad. Sci. Paris}, 226(216), 1948.

\bibitem{witten}
E.~Witten.
\newblock A note on {E}instein, {B}ergmann, and the fifth dimension.
\newblock Available in \href{https://arxiv.org/abs/1401.8048}{arXiv:1401.8048}.

\end{thebibliography}

\end{document}